\documentclass[]{myspie}  

\usepackage[]{graphicx}
\usepackage[]{epsfig}
\usepackage[]{picinpar}
\usepackage[]{pslatex}

\newcommand{\HST}{{\em HST}\/}

\newcommand{\FUSE}{{\em FUSE}\/}
\newcommand{\GALEX}{{\em GALEX}\/}
\newcommand{\Galex}{{\em GALEX}\/}
\newcommand{\galex}{{\em GALEX}\/}

\newcommand{\lya}{Ly$\alpha$}
\newcommand{\lyb}{Ly$\beta$}
\newcommand{\ars}{$^{\prime\prime}$}

\newcommand{\oigs}{\:{\rm ergs\:cm^{-2}\:s^{-1}\:\AA^{-1}}}
\def\deg{\hbox{$^\circ$}}
\def\pcaption#1{\small\baselineskip=11pt{#1}}

\title{FORTIS: Pathfinder to the Lyman Continuum} 

\author{Stephan R. McCandliss, Kevin France, Paul D. Feldman, Karl Glazebrook,
Gerhardt Meurer,  Luciana Bianchi, H. Warren Moos, Jeffrey W. Kruk, William P.
Blair, and Ivan Baldry \supit{a} 
\skiplinehalf
\supit{a}The Johns Hopkins University, Department of Physics and Astronomy, 3400 North Charles Street, Baltimore, MD 21218, USA \\
}

\authorinfo{Further author information: \\S.R.M.: E-mail: stephan@pha.jhu.edu, Telephone: 1 410 516 5272}

\begin{document} 
  \maketitle 

\begin{abstract}
Shull et al.\cite{Shull:1999} have asserted that the contribution of stars, relative to quasars, to the metagalactic
background radiation that ionizes most of the baryons in the universe
remains almost completely unknown at all epochs. The potential to
directly quantify this contribution at low redshift has recently become
possible with the identification by \GALEX\ of large numbers of
sparsely distributed faint ultraviolet galaxies.  Neither STIS nor
\FUSE\ nor \GALEX\ have the ability to efficiently survey these sparse
fields and directly measure the Lyman continuum radiation that may leak
into the low redshift (z $<$ 0.4) intergalactic medium.  We present
here a design for a new type of far ultraviolet spectrograph, one that
is more sensitive, covers wider fields, and can provide spectra and
images of a large number of objects simultaneously, called the
Far-ultraviolet Off Rowland-circle Telescope for Imaging and
Spectroscopy (FORTIS).  We intend to use a sounding rocket flight to
validate the new instrument with a simple long-slit observation of the
starburst populations in the galaxy M83.  If however, the long-slit
were replaced with microshutter array, this design could isolate the
chains of blue galaxies found by \GALEX\ over an $\approx$~30' diameter
field-of-view and directly address the Lyman continuum problem in a
long duration orbital mission.  Thus, our development of the sounding
rocket instrument is a pathfinder to a new wide field spectroscopic
technology for enabling the potential discovery of the long
hypothesized but elusive Lyman continuum radiation that is thought to leak from low redshift galaxies and contribute to the ionization of the universe.
\end{abstract}



\section{INTRODUCTION}
\label{sect:intro}  

The {\it Galaxy Evolution Explorer} (\GALEX) has recently acquired
some very exciting ultraviolet imagery. These broad band images,
centered on 1516~\AA\ and 2267~\AA,  locate the brightest ultraviolet
objects in the sky.   The starburst regions in nearby grand design
spirals are revealed to be immediately adjacent to clumpy dust lanes
seen in optical images.  A large number of faint blue galaxies with
angular separations $\sim$~few arcminutes are in the process of being cataloged.
Questions concerning the dust and molecular content, chemical
enrichment,  mass loss rates, supernovae rates, star formation
histories, and the escape fraction of ionizing radiation from the
underlying stellar populations naturally arise.
This last question is of particular importance to the question of the
how the universe came to be ionized. 

We propose to build an instrument capable of searching directly for the escaping
fraction of ionizing radiation from star-forming galaxies below their
rest frame Lyman edge at 911.7~\AA\ and assess their contribution
to the metagalactic ionizing radiation field.  These observations 
require wide-field far ultraviolet spectroscopy at moderate resolving
power ($\sim$~1500), emphasizing the wavelength region below \lya.

Neither \galex\ nor the Space Telescope Imaging Spectrograph (STIS) on
board \HST, nor the {\it Far Ultraviolet Spectroscopic Explorer}
\FUSE\ have the ability to carry out an efficient spectroscopic survey
of these widely separated faint blue galaxy fields below \lya.
\Galex\ has only rudimentary slitless spectroscopic capability. Its low
spectral resolving power ($R$~$\sim$~200), spatial resolution
($\sim$~5\ars) and bandpass restricted to $>$ 1350~\AA, cannot provide
the spectroscopic detail from which kinematic, chemical, thermal, and
density diagnostics can be derived.  STIS has demonstrated the
multiplexing power of long-slit spectroscopy\cite{Sonneborn:1998,Feldman:2000}, but it too is limited to
wavelengths  $>$ 1150~\AA.  And while the primary mission of \FUSE\ is
high resolution spectroscopy below 1190 \AA, its fast optics and the
high resolving power produces large astigmatic images that are
unsuitable for imaging spectroscopy.  Further, the high resolving power
makes it extremely difficult to observe objects fainter than $\sim$
10$^{-15} \oigs$.

\vspace{-.25in}
\begin{figwindow}[1,r,{\hspace*{0.5in}
\epsfig{file=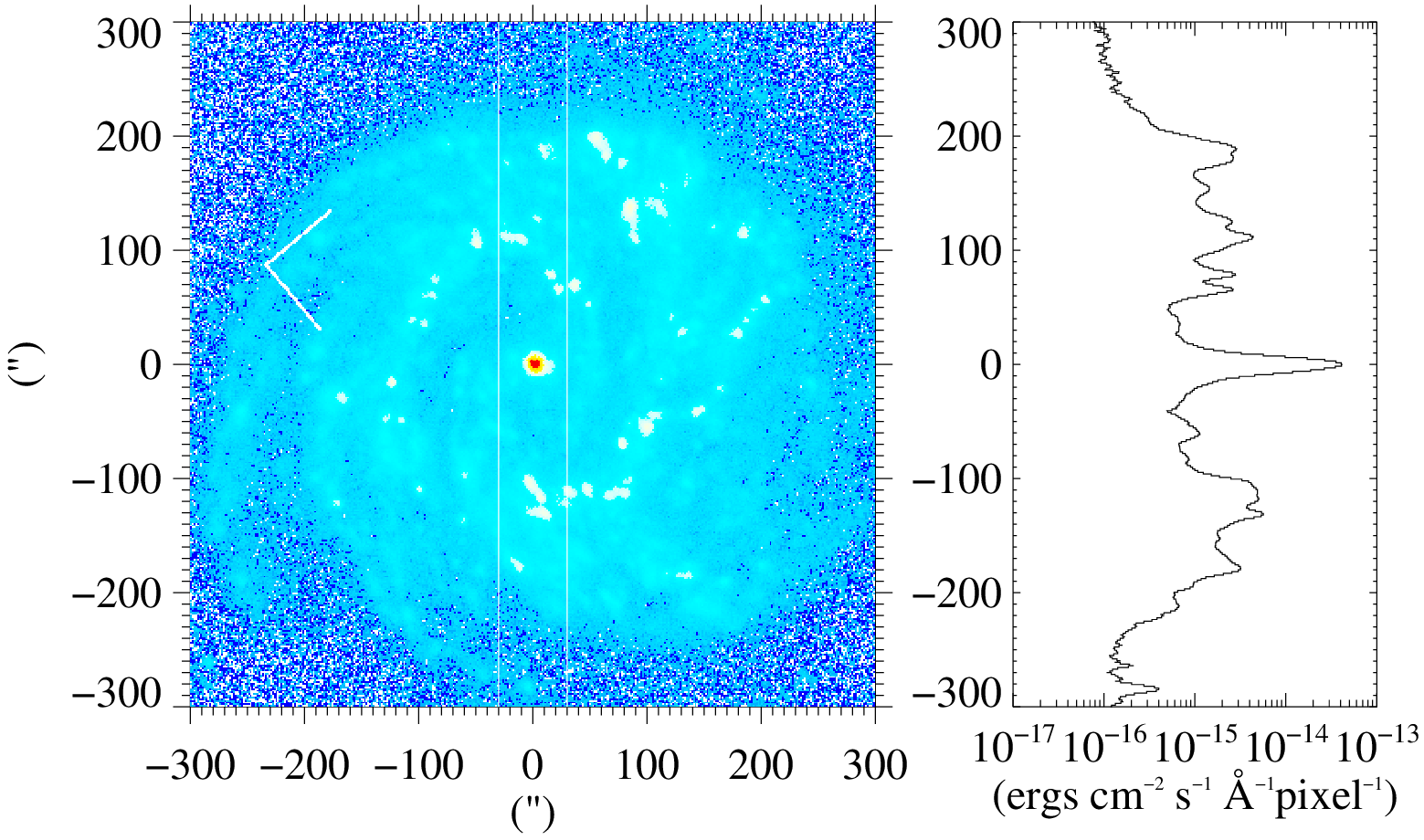,width=1.4in,angle=90} \vspace*{0.3in}},
\label{proj}
{\pcaption{\galex\ FUV image of M83 with 600\ars~$\times$~60\ars\ slit overlay.   The summed flux within the slit projected on the long axis is shown on the right. The flux scale on the abscissa is logarithmic. }}]

What is needed is a new approach to far ultraviolet spectroscopy, one
that is more sensitive, covers wider fields, and can observe a number
of objects simultaneously.  We present here the design of a highly
efficient wide field spectro/telescope based on a Gregorian telescope
with a ruled secondary, which we call the Far-ultraviolet Off
Rowland-circle Telescopy for Imaging and Spectroscopy (FORTIS).

We will briefly touch on some science drivers and our instrument
development plans in \S~\ref{science}.  We will then go on to discuss
far ultraviolet spectroscopic design evolution in \S~\ref{evol},
describe the instrument in \S~\ref{config} and present the expected
performance in \S~\ref{perform}.

\end{figwindow}

\section{Science Drivers and Development Plan}
\label{science}
It is widely thought that the ionizing background that permeates
metagalactic space is produced by quasars and star-forming galaxies.\cite{Madau:1996,Shull:1999,Heckman:2001}  Our ultimate
motivation is to address the assertion\cite{Shull:1999} that the
contribution of stars (relative to quasars) to the metagalactic
ionizing background remains almost completely unknown at all epochs.
The problem is to determine the average fraction of ionizing photons
$\langle f_{esc}\rangle$ produced by O and B stars that escape from
galaxies into the intergalactic medium (IGM).  This objective will
require a long duration observing program.

Our approach to this problem will be to use the
sounding rocket program to validate an advanced spectroscopic
instrument through a short duration observation with a focused science
objective.  This will in turn enable the rapid development of a mission
concept and associated technology that lie outside the capabilities of
existing orbiting observatories.

Our short duration observation will be of the nearby grand design
starburst galaxy M83. We will focus on measuring the
far ultraviolet spectral energy distributions of its three different
classes of hot star populations to probe the physics of star
formation.  We will demonstrate that this is a straightforward, albeit
technically challenging, observation for a 400 second sounding rocket
flight, by using the \galex\ observation of M83 as our guide.
The science requirements for the M83 observation will be used to derive
the instrument requirements for FORTIS.

Our long duration objective is to assess the galactic contribution to
the metagalactic ionizing radiation field by searching directly for the
ionizing radiation escaping from low redshift star-forming galaxies
below the rest frame Lyman edge at 911.7~\AA.
Heckman\cite{Heckman:2001} have pointed out that the modest sensitivity of
far ultraviolet telescopes to date (primarily HUT and \FUSE) has
limited the investigations to small and possibly unrepresentative
galaxy samples.
Deharveng\cite{Deharveng:1997} have argued that a quantitative assessment of
the contribution of galaxies to the ionizing background will require a
large number of observations before a Lyman continuum luminosity
function can be established.  Part of the problem, finding enough faint
ultraviolet galaxies, has just been solved by \galex, which is in the
process of cataloging large flux-limited samples of ultraviolet bright
galaxies.  A preliminary assay of the number of galaxies with Sloan
spectral identifications and redshifts z $<$ 0.25 in the FUV channel
with $m_{ab}~<~$22~ (fluxes $>$~ 7.6 $\times~10^{-17} \oigs$) finds 133
per square degree (Tamas Budavari, private communication).   We will show an
instrument derived from FORTIS will have the sensitivity to directly
search for Lyman continuum radiation escaping from a significant number
of these galaxies.  By replacing the long-slit of the new spectrograph
with a microshutter array\cite{Moseley:2000}, a unique and powerful
multi-object spectrograph is generated with the capability of recording
far ultraviolet spectra from large numbers of the widely separated very
blue galaxies.

Through our pathfinding M83 observation we will develop the core
technology required to search for the long hypothesized but elusive
Lyman continuum emission from low redshift galaxies.  This will allow
the important science question concerning the contribution of stars to
the metagalactic ionizing background to be quantified in a long
duration mission.  M83 is a well studied grand design spiral galaxy
with an extremely bright starburst core.  The \galex\ FUV image is
shown in Figure~\ref{proj}.  Surrounding the core are many bright blue
starburst clusters that trace the spiral arms.  The clusters are
coincident with \ion{H}{2} regions and are located in proximity to
complexes of dust and CO emission\cite{Rand:1999}.  Also apparent in
this image is a diffuse blue stellar component.  This diffuse stellar
component has been noted in other starburst
galaxies.\cite{Meurer:1995}   It fills the space between the bright
clusters and the spiral dust lanes and, despite the low surface
brightness, can contribute the dominant part of the total ultraviolet
luminosity of a typical starburst galaxy\cite{Tremonti:2001}. This
emission is not just dust scattered emission because in
\HST\ observations of nearby starburst galaxies the stars are
resolved.

\begin{figure*}[t]{\vspace{.25in}
\centerline{\hspace{.9in}\epsfig{file=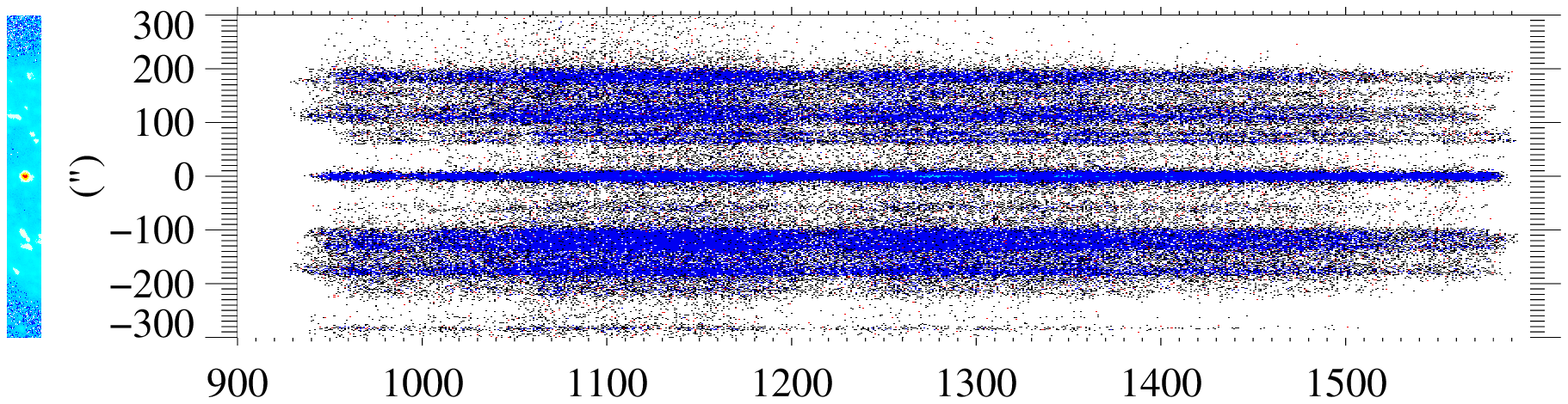,width=1.4in,angle=90}}\vspace{.2in}
\caption[]{\pcaption{Spectral simulation of the long-slit spectrum expected from M83.  The \galex\ observation was used to normalize the HUT starburst core spectrum at 1500 \AA.  The simulation was produced using the nominal \galex\ spatial resolution of 5\ars\ with 1.5\ars\ bins and the HUT spectral resolution of 3 \AA\ with 1 \AA\ bins.}}
\label{sim}}
\end{figure*}

The primary instrument requirement for observing M83 is to detect and spatially distinguish the three types of stellar populations within the 400 seconds of on target time available to a sounding rocket flight.    The HUT spectrum of the M83 core shows that the continuum flux at 1100~\AA\ is quite similar to that at 1500~\AA\ (3 $\times$~10$^{-13} \oigs$).  We use this observation along with the \galex\ flux calibrated FUV image of M83 at $\lambda_{eff}$ = 1516~\AA\ to simulate the emission of the various populations within each (1.5\ars)$^2$ square pixel, under the assumption that each pixel emits the same basic spectrum as the starburst core.  We include in the simulation a decreased dust attenuation with increasing distance from the core.

A slit 600\ars\ long will sample the entire diameter of M83.  We set the width 
to 60\ars\ to increase the number of compact clusters in the field of view
and to obtain good signal strength on the field population. 
The total mean flux at 1516 \AA\ in the region subtended by this slit in the
\galex\ FUV image at a position angle of (--48\deg\ from north) centered on the
starburst core  is 7.5~$\times$~10$^{-13} \oigs$.  
On the right side of Figure~\ref{proj}\ a profile of the flux along the long dimension  of the slit created by integrating over the short dimension is displayed.  There are $\approx$ 10 clusters some of which are at the resolution limit of \galex\ $\approx$ 5\ars.  Some of these clusters will undoubtedly fragment when viewed at higher spatial resolution and may reach the pointing accuracy limit of a typical sounding rocket $\sim$~1\ars. 

We produce the simulated spectra with 2~\AA\ (1~\AA\ bins), which is about a factor of 3 lower than our design goal.  The simulation of the long-slit spectrum is shown in Figure~\ref{sim}.  We used the effective area curve, discussed below, to convert the fluxes into total counts for an integration time of 400 seconds.  The count per bin arrival statistics have been simulated by adding or subtracting the square root of the counts per bin multiplied by a random number drawn from a normalized Gaussian distribution.  The expected count rate is $\approx$ 700 Hz.

FORTIS will be designed to have two pixel sampling at 1\ars.  The baseline plate scale is 33\ars\ mm$^{-1}$ and the inverse dispersion of the spectrograph is 20 \AA~mm$^{-1}$, so the conversion between wavelength and angle is 0.61 \AA~(\ars)$^{-1}$.  The 1\ars\ spatial resolution goal translates into a spectral resolution at 1000~\AA\ of $R$ = 1650, becoming somewhat higher at the longer wavelengths.   Our spectral resolution design goal is $\sim$~1500, which will allow molecular hydrogen content to be explored down to a level of $\sim$ 10$^{15}$ cm$^{-2}$ provided there is adequate signal to noise.   In practice this will be degraded by the finite extent of various stellar populations.  We may achieve the limiting spectral resolution on some of the tightest clusters.  The field populations are spread fairly uniformly between the clusters and will fill the slit yielding a slit limited spectral resolution of $R$ $\approx$ 30 at 1000\AA. Nevertheless, even at this low resolution, the integrated flux of the field population and any change in spectral slope can be determined.

The spectral bandpass for the instrument will span 900 -- 1600 \AA\ in order to provide a baseline to determine the  spectral slope at 1500~\AA\ to compare to the slope at 1100~\AA.  We will test whether the ratio of these two slopes is as  good a starburst age indicator as the ratio of the
\ion{C}{4}  $\lambda\lambda$ 1548 -- 1550 and \ion{Si}{4} $\lambda\lambda$ 1394 -- 1403 wind lines\cite{Leitherer:1999}.

\begin{figure*}
\centerline{\hspace*{-.2in}\epsfig{file=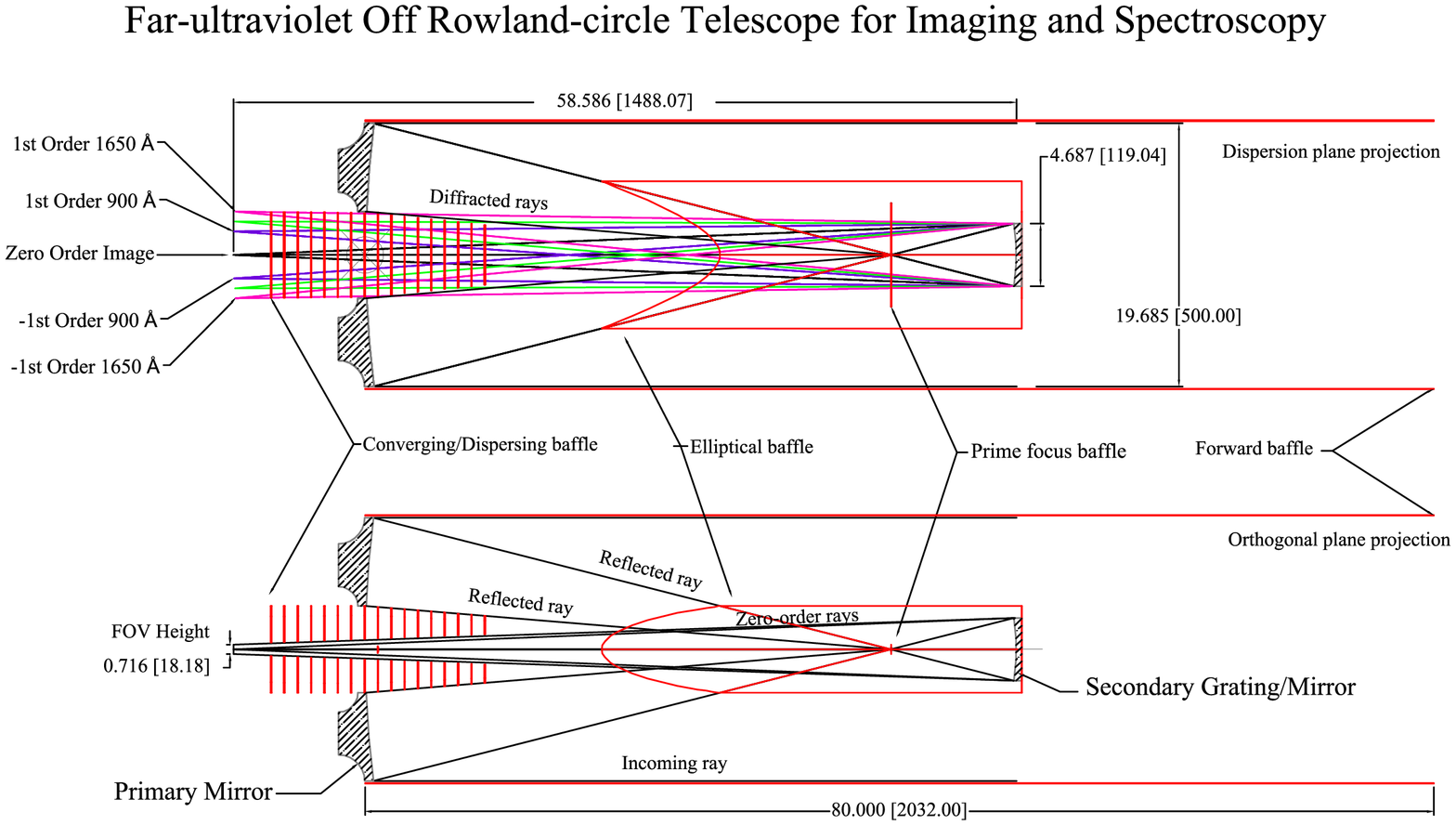,bbllx=1.5in,bblly=1.5in,bburx=7.25in,bbury=10in,height=6.in,angle=90}}
\caption[]{\pcaption{FORTIS sounding rocket optical layout.}}
\label{fortis}
\end{figure*}

\section{Far Ultraviolet Spectroscopic Design Evolution}
\label{evol}

The compelling new challenges to far ultraviolet science investigations emerging from the \galex\ surveys provide impetus to developing an instrument for acquiring spectra on a statistically significant number of very faint widely separated objects.  This is an especially difficult challenge in the far ultraviolet where the sources we wish to detect are weak ($<$~10$^{-15} \oigs$), geocoronal \lya\ is strong ($>$ 3 kRayleighs), and the efficiencies of the mirrors (reaching the Lyman limit) and detectors are low ($\sim$ 30 -- 40\%). 

Spectroscopic design in the far ultraviolet below \lya\ demands a minimalist approach. The ``two-bounce'' prime focus designs of the HUT, {\it OREFUS} and \FUSE\ observatories used relatively fast parabolic primary mirrors feeding the slit of a Rowland circle spectrograph.\footnote{In a Rowland circle spectrograph all the rays at wavelengths satisfying the grating equation entering a slit placed on a circle, with diameter is equal to the radius of the concave grating, are diffracted to a tangential focus on the same circle.} 
Although these two-bounce designs satisfy the requirement for efficiency they have poor off-axis spectral imaging performance.  This is because fast focal ratios tend to increase the height of the astigmatic image,  limiting the spatial resolution perpendicular to the dispersion.  Astigmatism control is a major challenge in Rowland circle spectrographs and methods have been developed to eliminate it at select wavelengths, either by controlling the grating figure\cite{Namioka:1961} or with holographic ruling methods\cite{Noda:1974}.  However, away from the corrected wavelengths astimgatism always grows and it grows slower when the focal length is long.

Long-slit spectroscopy in the far ultraviolet typically sacrifices efficiency for off-axis image quality.  Such designs  use ``three-bounce'' systems consisting of a  Cassegrain telescope feeding a Rowland circle spectrograph.  The  long effective focal length produces lower astigmatism, and can be used with long-slits out to several arcminutes off-axis.  Unfortunately, the efficiency sacrifice is high.  The third reflection drops the system effective area by a factor of 2.5 to 3, when SiC coated mirrors, which have (30 -- 40 \%) efficiency are used.  The low efficiency has caused us to develop a dual-order spectrograph that largely recovers the efficiency lost with the extra bounce.\cite{McCandliss:2003b}  

Noda\cite{Noda:1974} first wrote out in detail the general aberration
theory for holographic gratings, and since then techniques to minimize
spectroscopic aberrations have become increasingly sophisticated. They
reached a new level when solutions were found that made it possible to
control not only astigmatism but coma at the high ruling densities
required for the \FUSE\ mission \cite{Grange:1992}.  The high
resolution holographic gratings recorded for the Cosmic Origins
Spectrograph (COS -- slated to have been flown on the now canceled
servicing mission 4 to \HST) removed the spherical aberration with an
aspherical grating figure along with a \FUSE\ like aberration
minimization solution\cite{Green:2001}.  Recently developed numerically
optimized solutions for two-bounce systems achieve excellent narrow
band spectroscopic imaging with the grating used off the Rowland
circle\cite{Wilkinson:2001}.   An off-axis parabola design
\cite{Cunningham:2004} has a high enough spectral resolving power to
cleanly separate the \ion{O}{6}~$\lambda\lambda$1032 -- 1037 doublet
over a 0.5\deg\ field of view and yield 4 -- 9 \ars\ of spatial
resolution.
     
Our design evolves further with a very simple symmetric recording
geometry for producing a on-axis, normal-incidence dual-order
spectrograph, appropriate for moderate dispersion wide-field spectral
imaging at 1 -- 4\ars\ resolution.  Like all dual order
designs\cite{McCandliss:2003b} it uses a toric like grating surface to
correct for astigmatism in both positive and negative orders. Higher
order aberrations are corrected holographically.  There is a  zero
order field for imaging on the optic axis and 2 first order spectral
fields dispersed with mirror symmetry to either side of the optic
axis.
These fields are nearly contained within the central obscuration produced by  the secondary mirror/grating, in contrast to the off-axis parabola design of Cunningham\cite{Cunningham:2004}.  We have effectively a hybrid spectro/telescope that provides simultaneous imaging and spectroscopy.  We call our design the Far-ultraviolet Off Rowland-circle Telescope for Imaging and Spectroscopy (FORTIS).

\section{FORTIS Configuration}
\label{config}

\begin{table*}

\caption[]{\bf Instrument Summary \label{tab}}

\begin{tabular*}{6.75in}{ @{\extracolsep{\fill}}l@{\extracolsep{\fill}}l@{\extracolsep{\fill}}l@{\extracolsep{\fill}}}
\hline
Parameter  	&  FORTIS Sounding Rocket   &   FORTIS Long Duration   \\ \hline\hline
Secondary Focus Plate Scale 	& $33\!''$ mm$^{-1}$ & $33\!''$ mm$^{-1}$  \\
Primary Focus Plate Scale 	& $206.3\!''$ mm$^{-1}$ & $114.6\!''$ mm$^{-1}$  \\
Primary Focal Ratio	& f/2	& f/2	\\
System Focal Ratio	& f/12 & f/6.95 \\
Secondary Vertex to Focal Plane & 1488 cm & 2880 cm \\
Inverse Dispersion & 20 \AA\ mm$^{-1}$ & 20 \AA\ mm$^{-1}$ \\
Field of View	& 600\ars $\times$ 60\ars &  1800\ars $\times$ 1800\ars \\
Clear Area   &    1600 cm$^2$ & 5000 cm$^2$\\
Primary Diameter 	& 500 mm & 900 mm \\
Secondary Diameter 	& 120 mm & 400 mm\\
Bandpass		& 900 -- 1600 \AA\ &900 -- 1300 \AA\ \\
Spectral Resolution	& 1500	& 3000	\\
Spatial Resolution Perpendicular to Dispersion	& 1 -- 4\ars	& 0.5 -- 2\ars \\
Detector Pixel		& 0.015 $\mu$m	&  0.0075 $\mu$m \\
Approximate Detector Format	(1order)& 20 mm	$\times$ 40 mm & 60mm $\times$ 75mm \\
Effective Area Peak	& $\sim$ 50 cm$^2$ & $\sim$ 156 cm$^2$ \\
\hline

\end{tabular*}

\end{table*}

Here we review the baseline instrument design, covering the general optical path, efficiencies, detectors, baffling, raytrace and sensitivity.  The spectro/telescope configuration parameters are summarized in Table~\ref{tab}.

\paragraph{Optical Layout}
The telescope optical path is shown in Figure~\ref{fortis} in two
orthogonal views.  Light from an distance enters the telescope from the
right and reflects off the primary mirror converging to form an image
at the prime focus.  From the prime focus the light diverges to fill a
secondary grating.  The plane view (top of Figure~\ref{fortis}) shows how
the light is reflected into zero order and diffracted into the positive
and negative first orders of the grating. The reflected and diffracted
light comes to a secondary focus located behind the primary mirror.

The grating acts as a dichroic beam splitter. The relative spectral
bandpasses of the imaging and spectral regions are determined by a
combination of the groove efficiency and the reflectivity of the
grating.  A laminar shaped groove has an first order efficiency peak of
41\%  at a wavelength peak of $\lambda = 4h$, where $h$ is the groove depth  
 \cite{McCandliss:2001}.  The imaging efficiency is zero at
the 1st order peak wavelength and rises monotonically redward as the first order
efficiency falls (Figure~\ref{eff}).   Although it is possible to
develop sophisticated filter or even spectroscopic options to analyze
the zero order light more fully, such options go  beyond the scope of
this project.

The primary mirror and secondary grating will be coated with ion beam sputtered SiC with Al undercoating to keep the zero order reflectivity high \cite{McCandliss:1994}.  The diameter of the primary will be 500 mm, which
is set by the inner diameter of the largest sounding rocket skin that the NASA sounding rocket program provides.  We use an f/2 focal ratio for the primary, because slower beams require larger gratings, while faster beams cause the image quality to suffer.  We will construct a new optical bench for holding the primary, field stops, secondary, baffles and the (standard issue NASA provided) wide field star-tracker to common optical axes.  The optical bench will consist of an invar tube cantilevered off of a primary mirror baseplate.  The invar tube
also serves as a heat shield.

\paragraph{Baffles and Spectrograph Entrance Aperture}
Although it is possible to produce this type of instrument in a Cassegrain configuration, with a convex optic in front of the prime focus, a Gregorian configuration, with a concave optic behind the prime focus, has a distinct advantage.  It allows the inclusion of a field stop at the focus of the primary mirror to limit the field of view passed to the grating and avoids the image confusion in the spectral focal plane that would result from a slitless Cassegrain design. Most importantly it limits the solid angle for the geocoronal emission lines.  

The stop consists of a thin baffle with an aperture the width and
height of the 600\ars~$\times$~60~\ars\ slit.  The height of the baffle
can be made very small, because the prime focus plate scale is
de-magnified with respect to the secondary plate scale.  We match the
slit height to the detector height, so dispersed light from the grating
that comes from the region above or below the the slit does not enter
the detector.  The mismatch between the beam angles going to the
grating and departing the grating for the secondary focal plane allows
nearly all the light reflected or diffracted from the secondary to pass
by the baffle with minimal obscuration. 

Additional baffles are required to prevent the detectors from seeing
blank sky.  We have determined that the combination of a long tube the
diameter of the primary, an elliptically shaped central obscuration
extended to meet the converging beam from the primary and a series of
slotted plates in front and behind the primary are sufficent to
elliminate blank sky from the detector (see Figure~\ref{fortis}).

For the M83 observation a long-slit, will be etched in the prime focus
baffle plate.  For a Lyman continuum mission a microshutter array will
be integrated into the prime focus baffle plate.  The array is a grid
of apertures that can be either opened to let light pass or shut to
block light.  A cyrogenic version is currently in the development stage
for the near-infrared spectrograph (NIRSpec) on the James Webb Space
Telescope (JWST) \cite{Ge:2003}.  Typical dimension of the individual
apertures in these arrays are $\sim$ (0.1 mm)$^2$ \cite{Moseley:2000},
yielding an angular dimension of $\sim$(20\ars)$^2$ in the focal plane
of the primary.

\paragraph{Secondary Grating Figure and Recording Parameters}

The strong geocoronal \lya\ emission makes it important to use low scatter
optics.  Holographic gratings are the dispersion element of choice in
this bandpass as they are generally recognized to have $\sim$ 10 times
lower scattered light than  conventionally ruled blazed gratings
\cite{Flamand:1989,Mount:1978}.  The holographic etching process tends
to produce a groove with a symmetric profile, the consequence of which
is to yield equal power in both positive and negative orders
\cite{McCandliss:2001}.  We have chosen to exploit this well
documented property and increase our sensitivity by a factor of 2
through the use of two redundant detectors rather than experiment with blazing the groove of a holographic grating, as we have experienced
less than satisfactory results in the past.

A holographic grating is made by recording on a surface the
interference pattern of two coherent laser point sources placed at
($r_{\gamma}$, $\gamma$, $z_{\gamma}$) and ($r_{\delta}$, $\delta$,
$z_{\delta}$), as expressed in cylindrical coordinates with respect to
an origin at the grating normal.  We have found a holographic recording
solution for a normal incidence, ($\alpha$~=~0$^{\circ}$) off Rowland
circle grating that is particularly simple, $r_{\gamma} = r_{\delta}$ =
$S$, $\gamma = -\delta$,  and $z_{\gamma} = z_{\delta}$ = 0, where $S$
is the distance from the secondary vertex to the focal plane.  This
solution works because, $\alpha$~=~0$^{\circ}$,  the diffraction
angles are small, and the S is long.

One drawback of the dual order design is that the figure of the grating is required to have a different radius in the direction perpendicular to the dispersion to allow astigmatism to be corrected in both orders. \cite{McCandliss:2003b}  This is
accomplished with a rather modest reduction of the grating curvature in the
plane perpendicular to the dispersion following the prescription given by
\cite{Namioka:1961}.
We will perform a trade study to analyze whether this is the most practical configuration in terms of the extra expense of making a grating with a non-cylindically symmetric figure and the extra detector area required to cover both the spectral orders and the imaging order.

\section{FORTIS Expected Performance}
\label{perform}

\begin{figure*}[t]{\vspace*{.25in}
\centerline{\hspace*{-.4in}\epsfig{figure=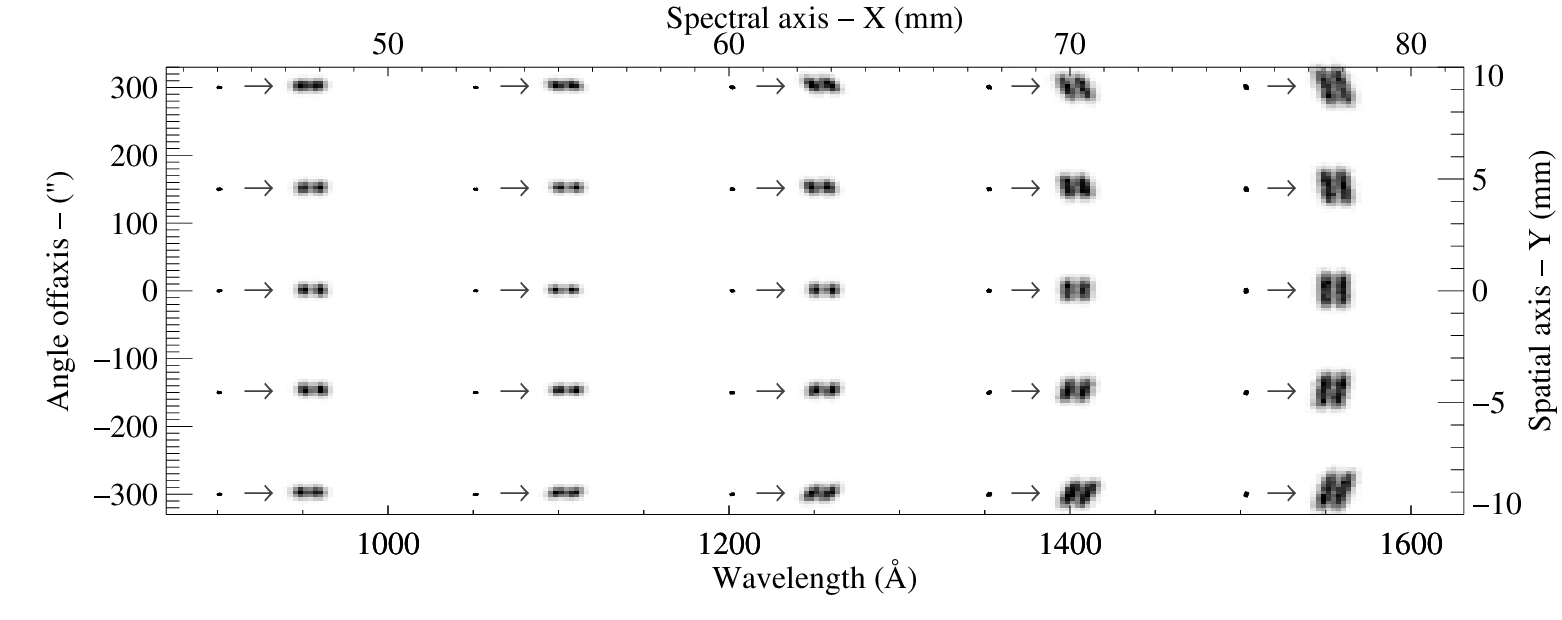,width=6.in}}
\caption[Raytrace]{\pcaption{Point spread functions as a function of wavelength and slit position calculated with a geometric raytrace.  Inset points are expanded by ten and sampled with (0.015 mm)$^2$ bins. Modest off-axis image rotation will be corrected with a distortion map.}}
\label{agpsf}}
\end{figure*}

\paragraph{Raytrace}

We have raytraced the design using the prescription given in Table~\ref{tab}.
The telescope primary and secondary figures (before astigmatism correction) are those appropriate to an aplanatic Gregorian \cite{Schroeder:1987}.
The input to the raytrace is a 5 $\times$ 5 array of Gaussian spots  spanning $\pm$300\ars and a wavelength range of  900 -- 1500 \AA.   The spatial width of each Gaussian spot is 1\ars\ full width half maximum (FWHM).  The spots come in wavelength pairs separated  by 1~\AA.  Ray bundles are generated from the spots and the resulting  distortions caused by the mirror surfaces and grating grooves
as the rays traverse the optical system are shown in Figure~\ref{agpsf}.
The degree of distortion depends on slit position and wavelength and have  mirror symmetry in $\pm$first order.  The spectral focal plane is convex as is expected from an aplanatic Gregorian.

The insets in Figure~\ref{agpsf} are the spots magnified by 10 and sampled with (0.015 mm)$^{2}$ bins.  On-axis (0\ars) near the astigmatism corrected wavelength of 1100~\AA\ there is very little 
distortion.  At longer wavelengths astigmatism increases but the spectral resolution is maintained.  The spatial resolution on-axis degrades at the longest wavelengths to $\approx$~4\ars.  Off-axis there is some rotation in the spot pairs, which can be corrected with an appropriate distortion map.
All spots show clean separation at 1~\AA\ indicating the spectral resolution exceeds this level at all wavelengths.  We find the Rayleigh resolution criteria is met in a projection of the on-axis point at 1050~\AA\ with a wavelength separation of 0.7~\AA\ yielding a spectral resolution of $R\approx$~1500. 

The zero order spots (not shown) exhibit a 4.6\ars\ FWMH astigmatism as expected from using a toric like secondary surface figure.  The width is 1\ars\ FWHM.  In the first year we will explore options to correct this image distortion by applying power to the calcium fluoride window in front of the imaging detector.

\begin{figure}[t]
\centerline{\psfig{figure=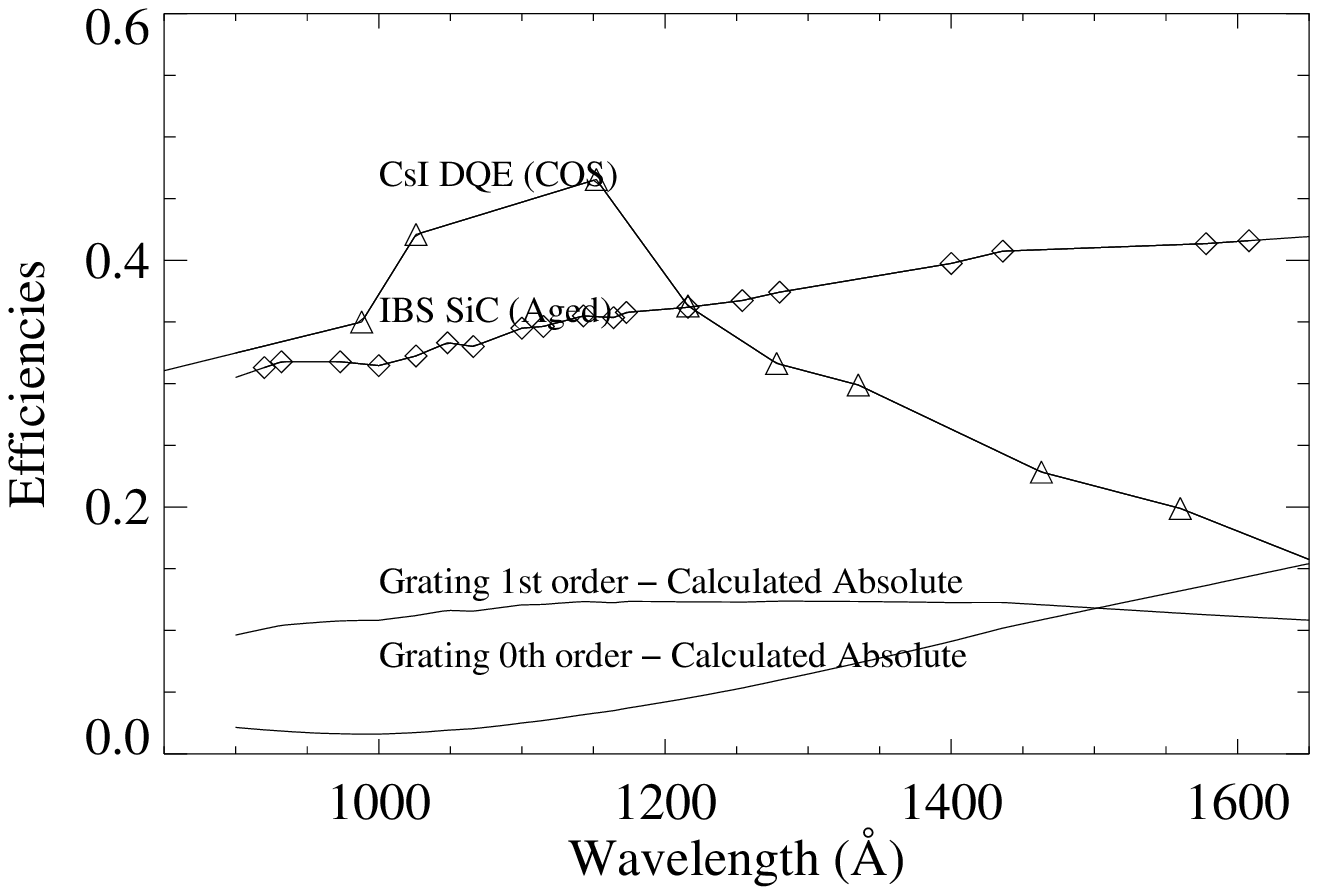,height=2.45in}
\hspace*{-.4in}
\psfig{figure=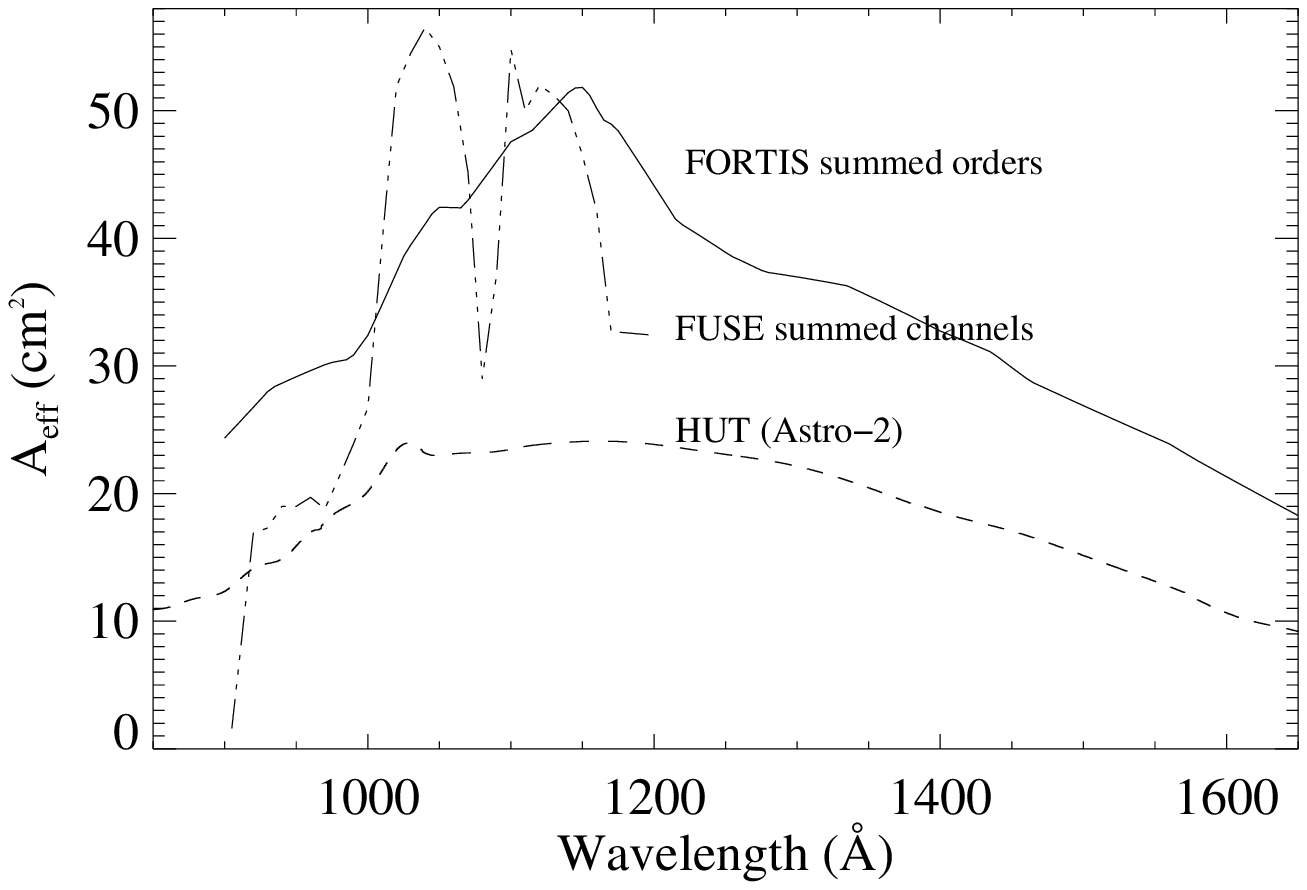,height=2.45in}
}
\vspace{-.2in}
\caption[]{\pcaption{Left, component efficiencies for estimating A$_{eff}$. Right, estimated A$_{eff}$ of FORTIS compared to \FUSE\ and HUT. FORTIS long duration is expected to have $\approx$ 3 times more area than FORTIS.}}
\label{eff}
\vspace*{-.2in}
\end{figure}

\vspace{-.1in}
\paragraph{Effective Area}
We calculate the effective with
\vspace{-.1in} \[ A_{eff}(\lambda) = A_{t}
R_{p}(\lambda) R_{sg}(\lambda) E_{g}(\lambda) 
Q_{d}(\lambda) T_{gr}, \]  where $A_{t}$ is the clear area of the telescope, $R_{p}$ and $R_{sg}$ are the reflectivities of the primary mirror and secondary grating,  $E_{g}$ is the groove efficiency of the grating, $Q_{d}$ is the quantum efficiency of the detector, and $T_{gr}$ is the transmission of the ion repeller grids.  Our component efficiencies are taken from \cite{McCandliss:2000}, which we have found to be in good agreement with the $A_{eff}$ of our current instrument.\cite{McCandliss:2003b}  The componet efficiencies are shown on the left in Figure~\ref{eff}.  On the right of this figure we show  the FORTIS effective area along with that for \FUSE\ \cite{Sahnow:2000}, and  HUT \cite{Kruk:1995}.

\vspace{-.1in}
\paragraph{Flux Detection Limits}

The ability to detect a faint source is limited by the detector background equivalent flux ({\it BEF}).  This is the signal that lies underneath all spectra that are extracted from the detector.  The {\it BEF} present in the 
spectral region of a point source changes as a function of wavelength due to
variations in the astigmatism height, the effective area, and scattered
light profile from geocoronal lines, mostly \lya.

\begin{figure*}[h]
\centerline{
\epsfig{figure=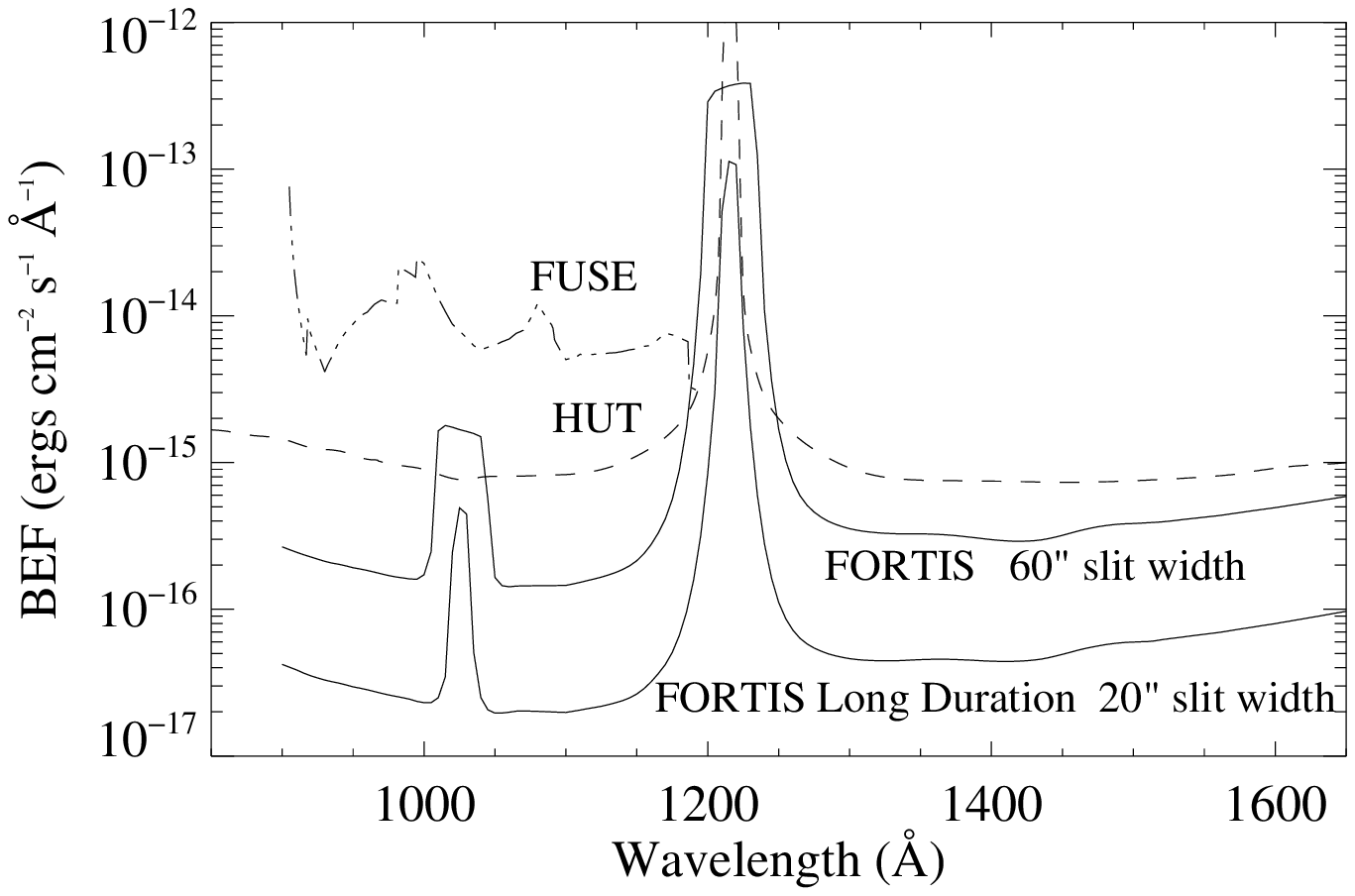,height=2.45in}
\hspace*{-.4in}
\epsfig{figure=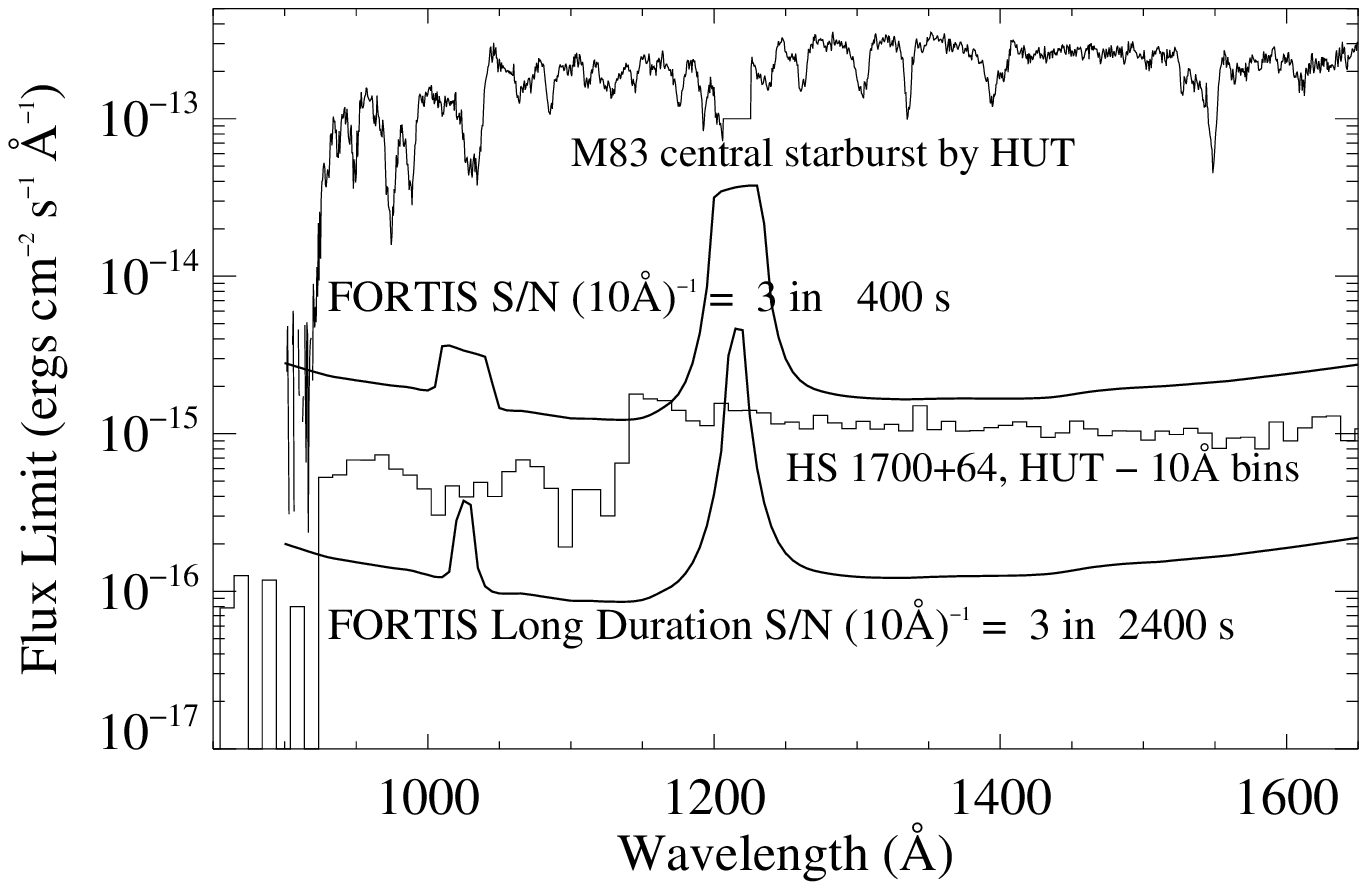,height=2.45in}
} 
\vspace{-.2in}
\caption[]{
\pcaption{Left,   background equivalent fluxes of FORTIS and FORTIS long duration
compared to \FUSE\ and HUT. Right,  3 $\sigma$ flux detection limit for 10~\AA\ bins in 400 s for FORTIS  and 2400 s for FORTIS long duration assuming the background equivalent fluxes. }}
\label{sens}
\end{figure*}

We estimate the {\it BEF} in ergs cm$^{-2}$ s$^{-1}$ \AA$^{-1}$   from
\vspace{-.05in} \[F_{\lambda} =\frac{h c}{\lambda}
\frac{B}{A_{eff}(\lambda)} \frac{H_{ast}(\lambda)}{D} \] where
$H_{ast}(\lambda)$ is the astigmatism height variation with wavelength
for our DOS, the \FUSE, and HUT spectrographs, $B$ is the background
rate, and $D$ is the dispersion, which is 40, 20, and 1 \AA\ mm$^{-1}$
for  HUT, FORTIS, and \FUSE\ respectively.  The total background rate
$B$ is a sum of the on-orbit dark rate, typically  1 Hz~${\rm
cm^{-2}}$ from radiation-induced background and intrinsic detector dark
counts, and the contribution of geocoronal \lya\ scattered from the
grating.  We estimate the wavelength variation of 3 kRayleighs of
\lya\ airglow from a 60\ars\ slit width using laboratory
measurements of a scatter profile from a typical holographic grating.
\lyb\ is also included at the level of 15 Rayleighs.
The result is shown on the left of Figure~\ref{sens}.  We  also show
an estimate for a FORTIS long duration instrument with a  20\ars\ slit width expected from a 
microshutter array.

The expected flux limits yielding a signal-to-noise per
10~\AA\ bin of 3 in 400 seconds for FORTIS  and 2400 seconds for FORTIS long duration are shown on the right
of Figure~\ref{sens}.  Overplotted are the HUT spectra of the M83
starburst core and the \ion{He}{2} Gunn-Peterson object HS 1700+64.
FORTIS will be the most sensitive far ultraviolet spectrograph
ever assembled.

\acknowledgments     

We would like to thank Erik Wilkinson for useful discussions regarding
off Rowland-circle grating techniques.  This work is supported through
NASA grant NNG04WC03G to JHU.


\bibliography{glas04}   
\bibliographystyle{spiebib}   

\end{document}